\documentclass[twocolumn]{fairmeta}
\newcommand{\algname}{\textbf{Kunlun}}
\usepackage{appendix}
\usepackage{makecell}
\usepackage{enumitem}
\setlist[itemize]{leftmargin=10pt}
\usepackage{amsmath}
\usepackage{amsfonts}
\usepackage{balance}

\usepackage{multirow}
\usepackage{algorithm}
\usepackage{algorithmic}
\usepackage{color,xcolor}

\definecolor{jackiecolor}{RGB}{220,120,170}

\title{Kunlun: Establishing Scaling Laws for Massive-Scale Recommendation Systems through Unified Architecture Design}

\author[*,1]{Bojian Hou}
\author[*,1]{Xiaolong Liu}
\author[*,\dagger,1]{Xiaoyi Liu}
\author[*,\dagger,1]{Jiaqi Xu}
\author[*,1]{Yasmine Badr}
\author[*,1]{Mengyue Hang}
\author[1]{Sudhanshu Chanpuriya}
\author[1]{Junqing Zhou}
\author[1]{Yuhang Yang}
\author[1]{Han Xu}
\author[1]{Qiuling Suo}
\author[1]{Laming Chen}
\author[1]{Yuxi Hu}
\author[1]{Jiasheng Zhang}
\author[1]{Huaqing Xiong}
\author[1]{Yuzhen Huang}
\author[1]{Chao Chen}
\author[1]{Yue Dong}
\author[1]{Yi Yang}
\author[1]{Shuo Chang}
\author[1]{Xiaorui Gan}
\author[1]{Wenlin Chen}
\author[1]{Santanu Kolay}
\author[1]{Darren Liu}
\author[\ddagger,2]{Jade Nie}
\author[\ddagger,2]{Chunzhi Yang}
\author[1]{Ellie Wen}
\author[1]{Jiyan Yang}
\author[1]{Huayu Li}

\affiliation[1]{Meta Platforms, Inc., Menlo Park, CA, USA.}
\affiliation[2]{OpenAI, San Francisco, CA, USA.}

\contribution[*]{Equal contribution}
\contribution[\dagger]{Corresponding author}
\contribution[\ddagger]{Work done while at Meta}

\abstract{Deriving predictable scaling laws that govern the relationship between model performance and computational investment is crucial for designing and allocating resources in massive-scale recommendation systems. While such laws are established for large language models, they remain challenging for recommendation systems, especially those processing both user history and context features. We identify poor \emph{scaling efficiency} as the main barrier to predictable power-law scaling, stemming from inefficient modules with low Model FLOPs Utilization (MFU) and suboptimal resource allocation. We introduce \algname, a scalable architecture that systematically improves model efficiency and resource allocation. Our low-level optimizations include Generalized Dot-Product Attention (GDPA), Hierarchical Seed Pooling (HSP), and Sliding Window Attention. Our high-level innovations feature Computation Skip (CompSkip) and Event-level Personalization. These advances increase MFU from 17\% to 37\% on NVIDIA B200 GPUs\footnote{\url{https://www.nvidia.com/en-us/data-center/b200/}} and double scaling efficiency over state-of-the-art methods. \algname\ is now deployed in major Meta Ads models, delivering significant production impact.}

\correspondence{\{xiaoyliu, jackiexu0313\}@meta.com}

\begin{document}

\maketitle

\section{Introduction}\label{sec:intro}

Click-through rate (CTR) prediction serves as the cornerstone of modern recommender systems, directly influencing both company revenue and user experience~\citep{guo2017deepfm,cheng2016wide,zhou2018din}. As these systems grow in scale and complexity, establishing predictable scaling laws---mathematical relationships describing how model performance improves with computational resources---has become increasingly critical for guiding architectural decisions and resource allocation~\citep{kaplan2020scaling,hoffmann2022training}. While scaling laws are well-established for large language models~\citep{kaplan2020scaling,brown2020language} and have been explored for non-sequential recommendation models~\citep{zhang2024wukong}, \textit{they remain an open challenge for systems that jointly model both sequential user behaviors and non-sequential context features}---a critical requirement for modern production recommendation systems at massive scale.

The success of scaling laws in language models suggests a natural approach: focus exclusively on sequential user behaviors. However, modern production recommender systems cannot adopt this simplified architecture. Non-sequential context features---including user demographics, ad metadata, and contextual signals---remain essential due to business requirements and deep infrastructure dependencies~\citep{han2025mtgr}. Any scalable architecture must therefore efficiently handle both feature types simultaneously, yet existing systems that attempt this joint modeling fail to achieve predictable power-law scaling behaviors.

We identify poor \textit{scaling efficiency}---the rate at which performance improves per unit of computational investment---as the primary barrier preventing predictable scaling laws. Specifically, scaling efficiency requires simultaneously optimizing both algorithmic effectiveness (performance gain per FLOP) and computational efficiency (hardware utilization). When examining state-of-the-art model paradigms that incorporate both sequence and non-sequence context features~\citep{zeng2024interformer}, we observe limited scaling efficiency due to two critical bottlenecks:
\textit{(1) Inefficient modules}: Models achieve only 3-15\% Model FLOPs Utilization (MFU), compared to 40-60\% for LLMs, due to heterogeneous feature spaces resulting in small embedding dimensions, irregular tensor shapes, and memory-bound operations.
\textit{(2) Inefficient computation resource allocation}: Naively scaling all components equally leads to diminishing returns, as different layers and event types benefit from different computational patterns.

\begin{figure*}
\centering
\includegraphics[width=.9\linewidth]{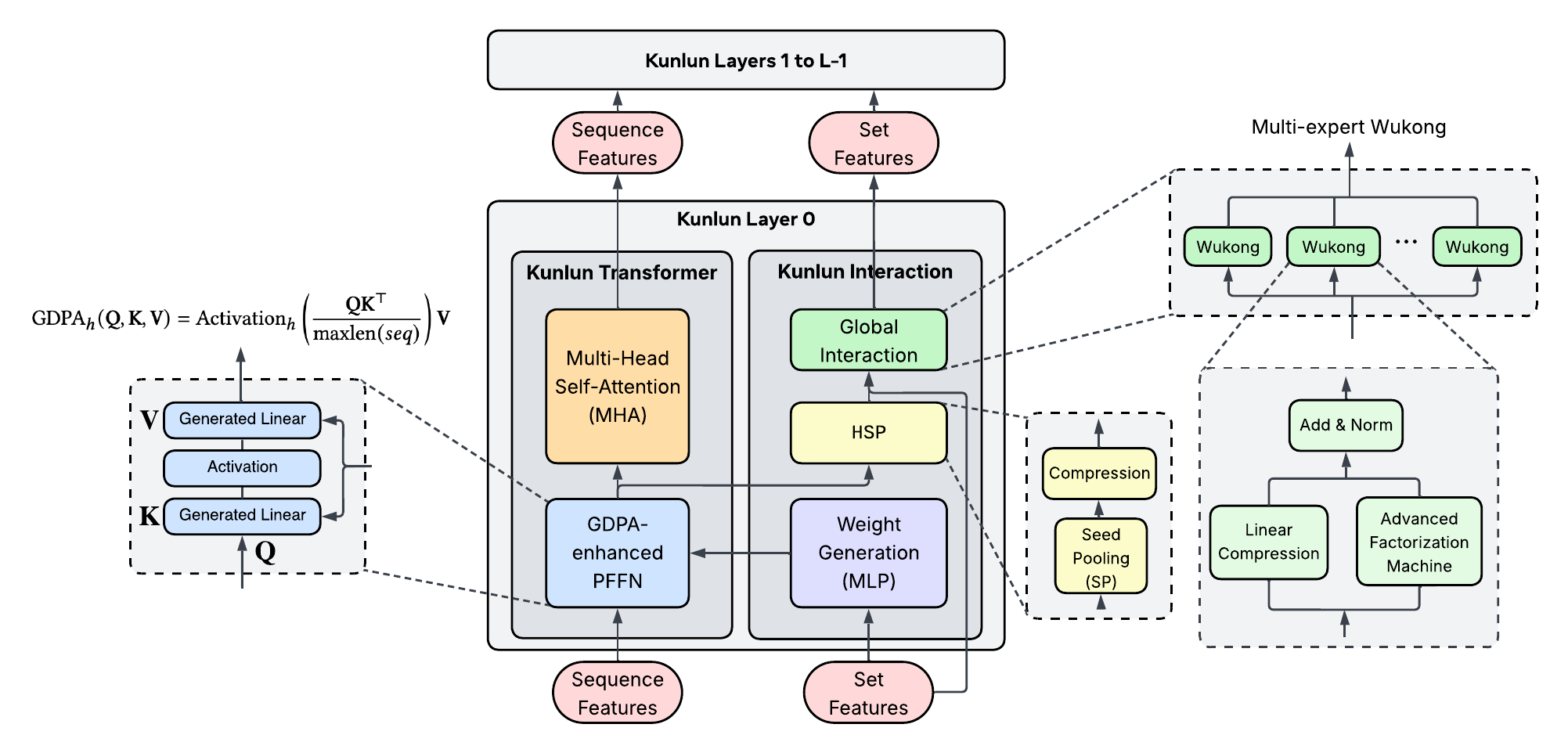}
\vspace{-0.4cm}
\caption{\label{fig:kunlun} Overview of the \algname\ architecture. The model is composed of multiple stacked layers, and each layer includes two main components: (1) a Kunlun Transformer block, which incorporates GDPA-enhanced PFFN and Multi-Head Self-Attention (MHA) to enable context-aware sequence modeling; and (2) a Kunlun Interaction block, which contains a Weight Generation module (to derive personalized weights for the PFFN from non-sequential features), a HSP module (to efficiently summarize sequential information for subsequent global interaction), and a Global Interaction module that facilitates interactions between sequential and non-sequential inputs, as well as interactions within the non-sequential features themselves.}
\end{figure*}

\noindent\textbf{Our Approach: Model-Efficiency Codesign.} We propose \algname\ (named after the Kunlun Mountains, symbolizing a foundational peak that unifies diverse elements), a unified architecture that tackles these challenges through systematic codesign at two levels. As illustrated in Figure~\ref{fig:kunlun}, \algname\ adopts a multi-layer architecture where each layer processes both sequence and non-sequence features through two main components: (1) a \textit{Kunlun Transformer Block} for context-aware sequence modeling via GDPA-enhanced personalized feedforward networks and multi-head self-attention, and (2) a \textit{Kunlun Interaction Block} for bidirectional information exchange through personalized weight generation, hierarchical sequence summarization, and global feature interaction.

\textit{Low-level module optimization} improves computational efficiency of fundamental building blocks. We introduce Generalized Dot-Product Attention (GDPA) for efficient personalized sequence transformation, Hierarchical Seed Pooling (HSP) for effective sequence summarization, and Sliding Window Attention for linear-complexity sequence modeling.

\textit{High-level computation reallocation} enables adaptive resource distribution. We propose Computation Skip (CompSkip) for layer-wise component selection and Event-level Personalization for importance-based resource allocation across heterogeneous event types.

Through this codesign approach, \algname\ improves MFU from 17\% to 37\% on NVIDIA B200 GPUs and achieves \textbf{2$\times$ scaling efficiency improvement over state-of-the-art approaches}, enabling the first predictable scaling laws for joint sequence-nonsequence modeling in recommendation systems.

\textbf{Contributions.} We summarize our main contributions as:
\begin{enumerate}[itemsep=0pt,parsep=0pt]
\item \textbf{Problem Formulation:} We identify and formalize the scaling efficiency challenge in massive-scale recommendation systems, highlighting two key bottlenecks: inefficient modules and inefficient computation resource allocation.
\item \textbf{Architecture Design:} We propose \algname, featuring systematic model-efficiency codesign with low-level optimizations (GDPA, HSP, Sliding Window Attention) and high-level innovations (CompSkip, Event-level personalization).
\item \textbf{Scaling Laws:} We demonstrate predictable scaling behavior, achieving 2$\times$ scaling efficiency improvement over state-of-the-art and establishing the first scaling laws for joint sequence-nonsequence modeling at massive scale.
\item \textbf{Deployment Impact:} \algname\ has been deployed at major Meta Ads models, achieving 1.2\% improvement in topline metrics and significant online impact.
\end{enumerate}

\section{Related Work}\label{sec:related}

\subsection{Recommendation System Architectures}

\textbf{Non-Sequential Methods.} Traditional recommendation models focus on learning interactions among non-sequential features via various mechanisms. Factorization Machines (FM)~\citep{rendle2010factorization} model pairwise feature interactions, while deep learning approaches combine FMs with neural networks~\citep{lian2018xdeepfm,wang2021dcn} for higher-order interactions. Wukong~\citep{zhang2024wukong} demonstrates scaling properties through the stacking of interaction modules; however, sequence modeling is not a central aspect of this architecture
~\citep{zhu2025rankmixer}.

\noindent\textbf{Sequential Methods.} Sequential recommendation models~\citep{zhou2018din,zhou2019dien} leverage RNNs or attention mechanisms to capture temporal dynamics in user behaviors. HSTU~\citep{zhai2024actions} employs transformer-style architectures for sequential recommendation, but primarily focuses on sequence-only modeling and does not effectively handle scenarios where rich non-sequential context features are present and powerful~\citep{zhang2025onetrans,chai2025longer,xu2025climber}. Recently, Interformer~\citep{zeng2024interformer} proposed an interleaving structure to address the limitation of unidirectional information flow through learning of bidirectional interactions between sequences and non-sequence context features. Although it demonstrates strong performance, its scaling efficiency at massive scales is limited by computational bottlenecks, which our work addresses.

\subsection{Scaling Laws in Machine Learning}

Scaling laws describe how model performance improves with increased computation, data, or parameters~\citep{kaplan2020scaling}. While well-established for language models, scaling laws for recommendation systems remain underexplored, particularly for models jointly learning sequential and non-sequential context features. Recent work~\citep{zhang2024wukong} demonstrates promising scaling trends but does not address computational efficiency or establish predictable scaling behaviors at massive scales.

\subsection{Model-Efficiency Co-design}


Efficient model design has gained increasing attention~\citep{dao2022flashattention,kitaev2020reformer}. Approaches include architectural innovations (e.g., sparse attention), system optimizations (e.g., operator fusion, mixed precision~\citep{micikevicius2018mixed}), and training techniques (e.g., gradient checkpointing). A complementary line of work improves serving efficiency by transferring a foundation model's capacity to compact models; for instance, LoopFM~\citep{jiang2026loopfm} structures a foundation model's historical intermediate embeddings as input features, going beyond scalar knowledge distillation. Our work stands out by systematically enhancing efficiency at both the low-level module and high-level computation allocation stages, specifically tailored to the unique characteristics and massive scale of recommendation systems.

\section{Preliminaries}\label{sec:prelim}

\textbf{Notation.} We use bold uppercase letters for matrices (e.g., $\mathbf{X}$), bold lowercase letters for vectors (e.g., $\mathbf{x}$), and lowercase letters for scalars (e.g., $n$). We use superscript $u$ to denote users, subscripts $i$ and $t$ to denote item and timestamp, respectively (e.g., $y^u_{i_t}$). We use $\mathbf{x}_j^{(l)}$ to denote the $j$-th non-sequence feature at the $l$-th layer, and $\mathbf{s}_t^{(l)}$ to denote the sequence feature at the $l$-th layer and timestamp $t$. We consider scenarios with $m$ dense features, $n$ sparse features, and $k$ sequences of length $T$. We use $d$ to denote the embedding dimension. Superscript $(l)$ denotes layer index, where $(0)$ represents raw input and $(1)$ represents the output after preprocessing.

\subsection{Problem Formulation: CTR Prediction with Heterogeneous Features}

Click-through rate (CTR) prediction estimates the probability of a user clicking on an item given heterogeneous information from multiple modalities. This heterogeneity is fundamental to modern recommendation systems, as user interests are depicted from different aspects: static context provides explicit preferences while behavior sequences reveal implicit dynamics.

Formally, given a user set $\mathcal{U}$ and an item set $\mathcal{I}$, the interaction sequence of user $u\in\mathcal{U}$ is defined as $S^u = [i^u_1,i^u_2,\dots, i^u_T]$, where $i^u_t \in \mathcal{I}$ is the item interacted at time step $t$ by user $u$. Our goal is to estimate the probability of user $u$ clicking on a new item $i_{T+1}$, denoted as $y^u_{i_{T+1}}$, given the historical interaction sequence $S^u$. Formally, we seek to learn a function $f: \mathcal{U} \times \mathcal{I}\times \mathcal{S} \rightarrow [0, 1]$, where $\mathcal{S}$ is the set of all possible sequences, such that:
\begin{equation}\label{eq:ctr}
P(y^u_{i_{T+1}} = 1 | u, i_{T+1}, S^u; \theta) = f(u, i_{T+1}, S^u; \theta).
\end{equation}

\subsection{Evaluation Metrics}\label{sec:eval_metrics}

We adopt \textbf{Normalized Entropy (NE)}~\citep{he2014practical} as our primary performance metric, defined as the ratio of model cross-entropy loss to background entropy (see Appendix~\ref{app:metrics} for detailed derivation). Lower NE indicates better performance, with $\Delta$NE $< 0$ meaning the model outperforms baseline prediction. We use NE over AUC because NE directly reflects calibration quality critical for production ranking systems.
For efficiency evaluation, we measure:
\begin{itemize}
\item \textit{Model FLOPs Utilization (MFU)}~\citep{DBLP:journals/corr/abs-2204-02311}: The ratio of achieved FLOPs to peak hardware FLOPs, indicating how effectively the implementation utilizes GPU compute resources.
\item \textit{Query Per Second (QPS)}: Training throughput that measures how many samples can be processed per second across the distributed training system.
\item \textit{GFLOPs}: Giga floating-point operations, where 1 GFLOPs = $10^9$ FLOPs. We report per-sample forward pass FLOPs to characterize model computational cost.
\end{itemize}

\subsection{Scaling Laws and Efficiency}
\label{sec:scaling_laws}

\textbf{Scaling Laws.} A scaling law describes the predictable mathematical relationship between model performance and computational resources. Following the neural scaling law literature~\citep{kaplan2020scaling}, we model this relationship as a power law in the logarithmic domain:
\begin{equation}\label{eq:scaling_law}
\text{NE}(C) = \text{NE}_0 - \eta \cdot \log(C / C_0)
\end{equation}
where $C$ represents the total compute investment (measured in FLOPs), $C_0$ is a baseline compute budget, $\text{NE}_0$ is the baseline performance, and $\eta > 0$ is the \textit{scaling coefficient} that determines the rate of NE improvement per unit of log-scaled compute. Larger $\eta$ indicates faster improvement---the NE decreases more rapidly as compute increases (i.e., steeper scaling curves). The validity of a scaling law requires two properties: (1) \textit{consistency} across different scales, meaning the functional form holds as we increase compute, and (2) \textit{predictability}, enabling reliable extrapolation to larger budgets.

\noindent\textbf{Scaling Efficiency.} While scaling laws describe \textit{whether} performance improves with compute, scaling efficiency quantifies \textit{how effectively} this improvement occurs. Taking the derivative of Equation~\ref{eq:scaling_law}, we see that $\eta = -d(\text{NE})/d(\log C)$ represents the slope of the NE vs. log(Compute) curve. We define scaling efficiency as this slope normalized by a historical baseline:
\begin{equation}\label{eq:scaling_eff}
\text{Scaling Efficiency} = \frac{\eta}{\eta_{\text{baseline}}} = \frac{\Delta \text{NE} / \log(C / C_0)}{\eta_{\text{baseline}}}
\end{equation}
where $\eta_{\text{baseline}}$ is the scaling coefficient of a baseline system (corresponding to 1X efficiency). For example, 2X scaling efficiency means the model achieves 2$\times$ the NE improvement per unit of log-scaled compute compared to baseline---visualized as a 2$\times$ steeper slope in the NE vs. log(Compute) curve.

Critically, scaling efficiency decomposes into two factors:
\begin{enumerate}
\item \textit{Algorithmic effectiveness}: How much performance gain (NE improvement) each FLOP provides. This depends on model architecture, training algorithms, and how well they exploit available compute for learning.
\item \textit{Computational efficiency}: What fraction of theoretical peak FLOPs are actually utilized by hardware, measured by Model FLOPs Utilization (MFU). Low MFU means the hardware idles while waiting for memory transfers or synchronization, wasting compute capacity.
\end{enumerate}

Therefore, achieving high scaling efficiency requires simultaneous optimization of both dimensions: designing algorithms that extract maximum learning signal per FLOP \textit{and} implementing them to maximize hardware utilization.

\section{Method}\label{sec:method}

In this section, we present \algname's architecture and its co-design in detail. We build upon the Interformer~\citep{zeng2024interformer} framework by introducing systematic model-efficiency co-design across two levels: \emph{low-level} module optimization for computational efficiency (Section~\ref{sec:low_level}) and \emph{high-level} computation reallocation for optimal resource distribution (Section~\ref{sec:high_level}).

\subsection{Architecture Overview}

\algname\ adopts a multi-layer architecture where each layer processes both sequence and non-sequence features through systematic interaction. As illustrated in Figure~\ref{fig:kunlun}, each layer is composed of two main components:

\noindent\textbf{(1) Kunlun Transformer Block} performs context-aware sequence modeling through two key mechanisms: (i) GDPA-enhanced Personalized FeedForward Networks (PFFN) that transform sequence embeddings using non-sequential context as personalized guidance, and (ii) Multi-Head Self-Attention (MHA) that captures dependencies within the sequential data itself.

\noindent\textbf{(2) Kunlun Interaction Block} facilitates bidirectional information exchange between sequences and non-sequences through three integrated modules: (i) a Weight Generation module that derives personalized weights for the PFFN from non-sequential features, (ii) a HSP (Hierarchical Seed Pooling) module that efficiently summarizes sequential information into compact representations for subsequent global interaction, and (iii) a Global Interaction module that models feature interactions between the sequence summaries and non-sequential features, as well as interactions within the non-sequential features themselves.

The model stacks $L$ such layers, where layer $l$ takes as input the outputs from layer $l-1$, enabling hierarchical feature learning at progressively higher levels of abstraction. This design enables \algname\ to generate NE gains through both vertical stacking (depth-wise feature refinement across multiple layers) and horizontal stacking (width-wise capacity expansion within each layer). Throughout all layers, we maintain the bidirectional information flow that enables mutually beneficial learning between different data modalities---non-sequential context guides sequence modeling, while sequence summarization informs non-sequence interaction.


\subsection{Feature Preprocessing}
Before the first layer, raw features are converted to unified embedding representations. The feature preprocessing follows standard pipelines~\citep{zeng2024interformer}, detailed in Appendix~\ref{app:preprocessing}, and we briefly summarize the key notations here.

\noindent\textbf{Non-Sequence Features.} Dense features are concatenated and projected: $\mathbf{x}_{\text{dense}}^{(1)} = \mathbf{W}_{\text{dense}}\mathbf{x}_{\text{dense}}^{(0)} \in \mathbb{R}^{d}$. Sparse features are embedded via lookup: $\mathbf{x}_{\text{sparse}_i}^{(1)} = \mathbf{W}_{\text{sparse}_i}\mathbf{x}_{\text{sparse}_i}^{(0)} \in \mathbb{R}^{d}$. All features are concatenated into the unified representation $\mathbf{X}^{(1)} \in \mathbb{R}^{(n+1) \times d}$.

\noindent\textbf{Sequence Features.} Multiple sequences from different event types are concatenated along the embedding dimension and fused via MLP into $\mathbf{S}^{(0)} \in \mathbb{R}^{T \times d}$.

\noindent\textbf{Rotary Temporal Embeddings (ROTE).} We extend Rotary Position Embeddings (RoPE)~\citep{su2024roformer} to incorporate temporal information. Unlike LLMs where token \textit{position} is the primary signal, recommendation sequences exhibit patterns where the \textit{temporal gap} between events carries crucial information---a click from yesterday differs fundamentally from one last month, even at adjacent positions. ROTE computes log-scaled timestamp gaps $\tau_t = \log(1 + \Delta t / \tau_{\text{scale}})$ and incorporates them into the rotary embeddings via $\mathbf{R}_{t,\tau_t} \mathbf{x}$, where the rotation matrix elements combine both positional frequencies $\theta_i$ and temporal frequencies $\phi_i$. This enables the model to capture both sequential order and temporal proximity. See Appendix~\ref{app:preprocessing} for detailed formulations.

\subsection{Low-Level Module Optimization}\label{sec:low_level}

To address computational inefficiency stemming from heterogeneous feature spaces, we redesign core modules to be compute-bound rather than memory-bound to improve MFU.

\subsubsection{Multi-Head PFFN with Generalized Dot-Product Attention (GDPA)}\label{sec:gdpa}

The Personalized FeedForward Network (PFFN) transforms sequence embeddings based on non-sequence context. Prior formulations~\citep{zeng2024interformer} express PFFN as:
\begin{equation}\label{eq:original_pffn}
\text{PFFN}(\mathbf{X}_{\text{sum}}^{(l)}, \mathbf{S}^{(l)}) = f(\mathbf{X}_{\text{sum}}^{(l)}) \mathbf{S}^{(l)},
\end{equation}
where $f(\mathbf{X}_{\text{sum}}^{(l)})$ is a two-layer MLP that generates transformation weights based on summarized non-sequence features $\mathbf{X}_{\text{sum}}^{(l)} \in \mathbb{R}^{n_{\text{sum}} \times d}$, and $\mathbf{S}^{(l)} \in \mathbb{R}^{T \times d}$ are the sequence embeddings at layer $l$.

\noindent\textbf{Computational Bottleneck.} This formulation suffers from low MFU due to small, memory-bound matrix operations with irregular shapes. Moreover, per-layer heavyweight activations and non-fusible back-to-back matrix multiplications further limit GPU utilization.

\begin{figure}[t]
    \centering
    \includegraphics[width=1\linewidth]{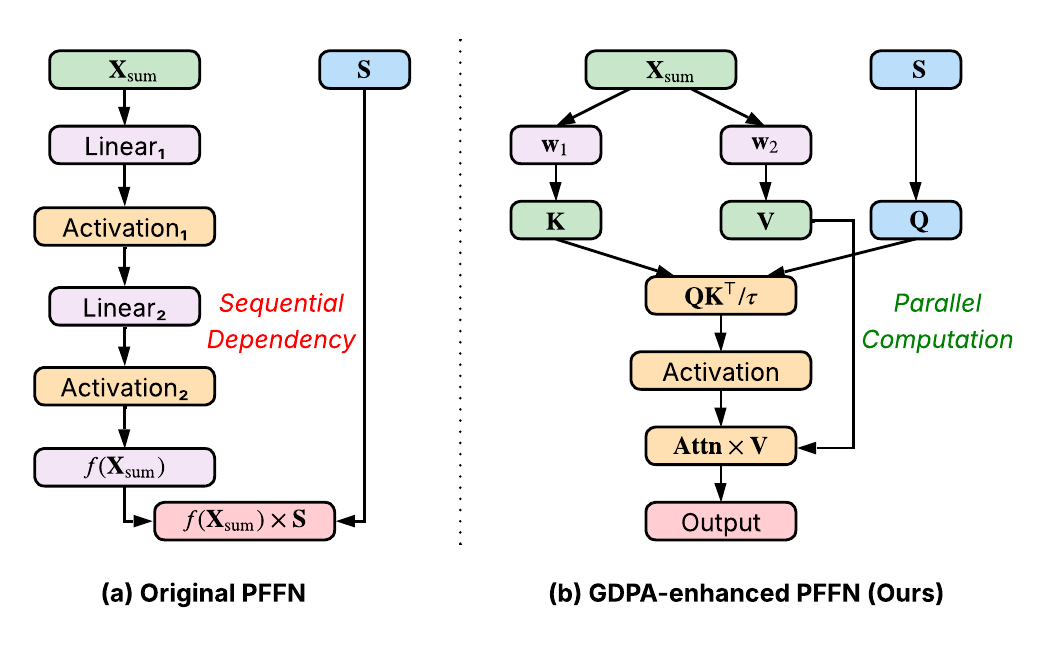}
    \vspace{-0.9cm}
    \caption{Comparison between (a) the original PFFN, and (b) our GDPA-enhanced PFFN. Note: Both are one-block demos.}
    \label{fig:comparison}
\end{figure}

\noindent\textbf{GDPA Reformulation.} To address these inefficiencies, we redesign PFFN as a multi-head attention-style operator, termed \emph{Generalized Dot-Product Attention} (GDPA), enabling the entire PFFN to be fused into a single kernel in a FlashAttention-like~\citep{dao2022flashattention, shah2024flashattention} manner. The key insight is that the transformation $f(\mathbf{X}_{\text{sum}}) \mathbf{S}$ can be viewed as a form of cross-attention where the input sequence serves as the query, while the two weight matrices from the personalized two-layer MLP $f(\mathbf{X}_{\text{sum}}^{(l)})$ function as the key and value, as depicted in the left-hand side of Figure~\ref{fig:kunlun}. Specifically, for each attention head $h \in \{1, \dots, H\}$, we compute:
\begin{equation}\label{eq:gdpa}
\text{GDPA}_h(\mathbf{Q},\mathbf{K},\mathbf{V})=\text{Activation}_h\left(\frac{\mathbf{Q}\mathbf{K}^\top}{\tau}\right)\mathbf{V},
\end{equation}
where $\mathbf{Q} = \mathbf{S}^{(l)}$ are the sequence tokens serving as queries, $\mathbf{K} = \mathbf{w}_1^{(h)}(\mathbf{X}_{\text{sum}}^{(l)})$ and $\mathbf{V} = \mathbf{w}_2^{(h)}(\mathbf{X}_{\text{sum}}^{(l)})$ are learnable projections from non-sequence context for head $h$, $\tau = \text{maxlen}(seq)$ is a temperature parameter for scaling ($\text{maxlen}(seq)$ works better than embedding dimension $d$ or $\sqrt{d}$ empirically), and Activation$_h$ is the activation function for head $h$. The outputs from all heads are then concatenated and projected through $\mathbf{W}^O$ to produce the final output. A comparison between PFFN, and GDPA is shown in Figure~\ref{fig:comparison}.

\noindent\textbf{Efficient Attention-Style Implementation.} Following Flash Attention style designs~\citep{dao2022flashattention, tillet2019triton}, we employ a memory-efficient, block-wise execution strategy that avoids materializing large intermediate tensors in the backward pass. The fused kernel is further optimized for PFFN workloads with small key/value shapes and jagged, variable-length inputs as discussed in Appendix~\ref{app:gdpa}, resulting in up to a 6$\times$ MFU improvement for the PFFN block.

\noindent\textbf{Residual Connection and PFFN Stacking.} Unlike prior PFFN formulations, GDPA incorporates a residual connection that enables stable PFFN stacking across multiple layers. This design allows the model to progressively refine sequence representations through deeper transformations, yielding additional NE gains that were not achievable with the original non-stackable PFFN design.

\subsubsection{Hierarchical Seed Pooling (HSP): Sequence Summarization}\label{sec:hsp}

Sequence summarization condenses long sequences ($T > 1000$) into compact representations for efficient non-sequence interaction. Traditional Pooling by Multihead Attention (PMA), as used in prior approaches~\citep{zeng2024interformer}, employs random learnable queries:
\begin{equation}\label{eq:pma}
\text{PMA}(\mathbf{S}) = \text{MHA}(\mathbf{Q}_{\text{learnable}}, \mathbf{S}, \mathbf{S}),
\end{equation}
where $\mathbf{Q}_{\text{learnable}} \in \mathbb{R}^{n_{\text{tokens}} \times d}$ are randomly initialized learnable queries and MHA refers to Multihead Attention~\citep{vaswani2017attention}.

\noindent\textbf{HSP Design.} We introduce Hierarchical Seed Pooling (HSP), which improves upon PMA through hierarchical compression using learnable seed embeddings. The key insight is that hierarchical summarization that first learns seed-level representations, then compresses them into final tokens provides better initialization and more stable training than direct random queries. Given input sequence $\mathbf{S} \in \mathbb{R}^{B \times T \times d}$, the HSP operation proceeds in three stages:

\textit{Stage 1: Seed Embedding Initialization.} Define learnable seed embeddings $\mathbf{E}_{\text{seed}} \in \mathbb{R}^{n_{\text{seeds}} \times d}$ where $n_{\text{seeds}} > n_{\text{tokens}}$ provides overcomplete representation. These seeds are shared across all samples in the batch.

\textit{Stage 2: Seed-level Attention.} Compute attention from seeds to sequence, producing seed-level representations:
\begin{equation}\label{eq:hsp_attn}
\mathbf{H}_{\text{seed}} = \text{MHA}(\text{Norm}(\mathbf{E}_{\text{seed}}), \mathbf{S}, \mathbf{S}) \in \mathbb{R}^{B \times n_{\text{seeds}} \times d},
\end{equation}
where $\text{Norm}(\cdot)$ applies LayerNorm or RMSNorm to the seed embeddings. Each batch sample now has $n_{\text{seeds}}$ seed-level representations aggregated from the sequence.

\textit{Stage 3: Parameter-Efficient Seed Compression (SumKronLinear).}
The third stage compresses the $S$ pooled seed vectors to $T$ output tokens. This is the critical bottleneck where compression quality directly impacts downstream model performance.

We introduce \textbf{SumKronLinear}, a Kronecker product-based compression layer that achieves parameter-efficient joint sequence-embedding transformation. Given input $\mathbf{X} \in \mathbb{R}^{B \times S \times D}$, the compression to $\mathbf{Y} \in \mathbb{R}^{B \times T \times D}$ is computed as:
\begin{equation}
\mathbf{Y}_b = \sum_{i=1}^{k} \mathbf{Z}_i^\top \mathbf{X}_b \mathbf{W}_i, \quad \text{where } \mathbf{Z}_i \in \mathbb{R}^{S \times T}, \mathbf{W}_i \in \mathbb{R}^{D \times D}
\end{equation}

This formulation provides two key advantages. First, \textbf{parameter efficiency}: the decomposition reduces parameters from $\mathcal{O}(S \cdot D \cdot T \cdot D)$ to $\mathcal{O}(k \cdot (S \cdot T + D^2))$. For typical configurations ($S{=}256$, $T{=}32$, $D{=}384$, $k{=}8$), this yields a $14\times$ reduction. Second, \textbf{cross-dimensional expressiveness}: unlike separable factorizations $\mathbf{Y} = \mathbf{P}_{\text{seq}}^\top \mathbf{X} \mathbf{P}_{\text{emb}}$ that assume independence between sequence and embedding dimensions (rank-1 structure), SumKronLinear captures joint correlations through rank-$k$ Kronecker structure. As $k$ increases, the representational capacity approaches that of a full linear transformation while maintaining sublinear parameter growth.

\noindent\textbf{Scaling Properties.} The parameter efficiency of SumKronLinear enables scaling the compression capacity without proportionally increasing model size. As embedding dimension $D$ increases, the relative parameter savings grow, making this approach increasingly favorable at massive scale. This enables flexible capacity allocation that can be tailored to match the characteristics of different workloads and event types.






\subsubsection{Sliding Window Attention}\label{sec:sliding_window}

Full self-attention in sequence modeling has quadratic complexity $O(T^2)$ where $T$ is sequence length. For sequences with $T > 1000$, this becomes prohibitive.

\noindent\textbf{Locality Bias in Recommendation.} Consistent with prior observations in sequential recommendation~\citep{zhou2018din,zhou2019dien}, recent interactions exhibit significantly higher predictive value than distant historical ones. This temporal decay in relevance motivates restricting attention to local windows.

\noindent\textbf{Windowed Pattern.} We apply sliding window attention with window size $w$:
\begin{equation}\label{eq:sliding_window}
\small
\text{Attention}(Q_t, K, V) = \text{softmax}\left(\frac{Q_t K_{[t-w:t+w]}^{\top}}{\sqrt{d}}\right) V_{[t-w:t+w]},
\end{equation}
where $Q_t$ is the query at position $t$, and $K_{[t-w:t+w]}, V_{[t-w:t+w]}$ are keys and values within the window $[t-w, t+w]$ at position $t$.

This reduces complexity from $O(T^2)$ to $O(Tw)$, achieving 31.1\% QPS improvement while preserving local dependencies critical for sequence modeling when $T > 1000$. The window size $w$ is configurable per event type through event-level personalization (Section~\ref{sec:event_personalization}).

\subsection{High-Level Computation Reallocation}\label{sec:high_level}

Beyond local efficiency, achieving high scaling efficiency requires intelligent resource distribution across layers and event types.

\subsubsection{Computation Skip (CompSkip)}\label{sec:compskip}

CompSkip selectively skips modules within layers via skip connections. Similar to layer interleaving strategies explored in recent LLMs~\citep{team2024gemma2} and findings on layer redundancy~\citep{men2024shortgpt}, we alternate computational patterns across layers in recommendation models.

\noindent\textbf{Every-Other-Layer Pattern.} We implement an alternating pattern where:
\begin{itemize}[leftmargin=*,itemsep=1pt,topsep=2pt]
\item \textit{Even layers (0, 2, 4, ...)}: Skip self-attention; compute fresh HSP summaries
\item \textit{Odd layers (1, 3, 5, ...)}: Skip HSP (reuse previous summaries) and PFFN; compute self-attention
\end{itemize}
This ensures each layer contributes either local refinement (self-attention) or global summarization (HSP), maintaining expressiveness while reducing computation. The detailed configuration algorithm is provided in Appendix~\ref{app:algorithm}.

\noindent\textbf{Efficiency Gains.} 
The every-other-layer pattern reduces FLOPs by $\sim$43.1\% and improves QPS by $\sim$35\% on average, maintaining model quality through complementary updates that preserve local and global sequence understanding.

\subsubsection{Event-Level Personalization}\label{sec:event_personalization}

Different event sequences have varying importance and characteristics. Clicks provide stronger signals than impressions, and long sequences may contain more noise than short ones. Event-level personalization allocates computational resources proportional to each sequence type's importance.

\noindent\textbf{Per-Event Configuration.} For each event type $e \in \{e_1, \dots, e_K\}$, we configure:
\begin{itemize}
\item $d_{\text{model}}^{(e)}$: Model dimension, controlling capacity
\item $n_{\text{heads}}^{(e)}$: Number of attention heads, controlling multi-faceted learning
\item $n_{\text{tokens}}^{(e)}$: Number of summarization tokens passed to non-sequence interaction
\item $L^{(e)}$: Number of processing layers, controlling depth
\item $w^{(e)}$: Sliding window size for self-attention, controlling receptive field
\end{itemize}

\noindent\textbf{Importance-Based Allocation.} High-value events (e.g., purchase conversions) receive larger $d_{\text{model}}$, more heads, more tokens, and more layers, while lower-value events (e.g., ad impressions) use smaller configurations. This enables efficient scaling across diverse behavior signals---allocating more compute to critical events while maintaining coverage of peripheral signals.

\noindent\textbf{Example Configuration.} For a system with click and impression sequences:
\begin{align}
\text{Click}: \quad & d_{\text{model}}=256, n_{\text{heads}}=8, n_{\text{tokens}}=32, \notag \\ &L=3, w=100 \notag \\
\text{Impression}: \quad & d_{\text{model}}=128, n_{\text{heads}}=4, n_{\text{tokens}}=16, \notag \\ &L=2, w=50 \notag
\end{align}

\subsubsection{Global Interaction Module with Mixture of Wukong Experts}\label{sec:global_interaction}

The Global Interaction module is a core component of the Kunlun Interaction Block that processes both summarized non-sequence features and sequence summaries from HSP to model complex feature interactions. Given non-sequence feature summaries $\mathbf{X}_{\text{sum}}^{(l)} \in \mathbb{R}^{B \times n_{\text{sum}} \times d}$ and sequence summaries $\mathbf{H}_{\text{summary}}^{(l)} \in \mathbb{R}^{B \times n_{\text{tokens}} \times d}$ at layer $l$, the Global Interaction module combines these representations and learns rich cross-modal interactions between sequential and non-sequential information.

\noindent\textbf{Mixture of Wukong Experts Architecture.} The Global Interaction module employs a mixture-of-experts architecture where multiple Wukong modules~\citep{zhang2024wukong} operate concurrently. Each Wukong module is a complete feature interaction model containing multiple complementary interaction functions: DOT product captures linear interactions and deep interaction networks learn hierarchical interactions. These functions within each Wukong module work together as an integrated unit to capture different interaction patterns (linear and hierarchical) rather than executing as independent parallel operations.

\noindent\textbf{Horizontal Scaling through Expert Parallelism.} Within each layer, the model employs $M$ Wukong experts that process different feature subsets concurrently. The combined input features $\mathbf{X}_{\text{global}}^{(l)} = \text{Concat}(\mathbf{X}_{\text{sum}}^{(l)}, \mathbf{H}_{\text{summary}}^{(l)}) \in \mathbb{R}^{B \times (n_{\text{sum}} + n_{\text{tokens}}) \times d}$ are partitioned across experts. Each expert $i \in \{1, \dots, M\}$ processes a designated feature partition:
\begin{equation}\label{eq:expert_parallel}
\mathbf{H}_i^{(l)} = \text{Wukong}_i(\mathbf{X}_{\text{global}, i}^{(l)}),
\end{equation}
where $\mathbf{X}_{\text{global}, i}^{(l)}$ represents the feature subset assigned to expert $i$. The expert outputs are then aggregated to produce the layer output. This expert parallelism enables horizontal scaling by distributing computation across multiple experts within a single layer, allowing the model to process wider feature representations efficiently.

\noindent\textbf{Vertical Scaling through Layer Stacking.} The model stacks $L$ layers vertically, where each layer refines representations learned from the previous layer:
\begin{equation}\label{eq:vertical_stacking}
\mathbf{X}_{\text{global}}^{(l+1)} = f^{(l)}(\mathbf{X}_{\text{global}}^{(l)}),~l \in \{0, \dots, L-1\},
\end{equation}
where $f^{(l)}$ represents the transformation at layer $l$ combining outputs from all $M$ experts. This vertical stacking enables the model to learn increasingly abstract feature representations through hierarchical refinement, with each layer building upon the interactions captured by previous layers.

Together, horizontal expert parallelism and vertical layer stacking enable \algname\ to generate NE gains through both width-wise capacity expansion (more experts per layer) and depth-wise feature refinement (more layers), providing flexible scaling along both dimensions to achieve optimal performance-efficiency trade-offs.

\subsection{Multi-Layer Architecture}\label{sec:multilayer}

\algname\ stacks $L$ layers, each containing a \textit{Kunlun Transformer Block} (GDPA-enhanced PFFN + Multi-Head Self-Attention for sequence modeling) and a \textit{Kunlun Interaction Block} (Weight Generation + HSP + Global Interaction for cross-modal learning). 

The architecture maintains \textbf{bidirectional information flow}~\citep{zeng2024interformer}: non-sequence features guide sequence modeling through personalized GDPA weights, while sequence summaries inform non-sequence interaction through the Global Interaction module. This enables mutually beneficial learning where both modalities progressively refine each other across layers.

Vertical stacking enables \textbf{hierarchical learning}---early layers capture basic patterns while deeper layers model high-order dependencies. The architecture demonstrates predictable \textbf{scaling properties}: NE improvement follows $\Delta \text{NE}_l \approx c / \log(l+1)$, validating power-law scaling along both depth and width dimensions.

After $L$ layers, the final non-sequence representation $\mathbf{X}^{(L)}$ from the Global Interaction module is fed to task-specific heads: $\hat{y} = \sigma(\text{MLP}(\mathbf{X}^{(L)}))$. See Appendix~\ref{app:multilayer} for detailed equations and information flow.

\section{Experiments}\label{sec:exp}

We conduct comprehensive experiments to evaluate \algname\ on large-scale internal datasets. Our experiments aim to answer the following research questions:
\begin{itemize}
\item \textbf{RQ1:} Does \algname\ achieve predictable and superior scalability for massive-scale recommendation systems?
\item \textbf{RQ2:} How do the low-level module optimizations (GDPA, HSP, Sliding Window Attention) contribute to efficiency improvements?
\item \textbf{RQ3:} How do the high-level computation reallocation strategies (CompSkip, Event-level Personalization) improve the model quality and efficiency?
\item \textbf{RQ4:} How does \algname\ perform in production deployment?
\end{itemize}

\subsection{Experimental Setup}

\textbf{Datasets.} We evaluate on large-scale internal datasets from Meta Ads: 70B+ samples with hundreds and thousands of non-sequence features and 10+ sequence features of length from hundreds to thousands, spanning multiple event types (clicks, conversions, impressions)

\noindent\textbf{Baseline Selection and Rationale.}
We compare against two representative production-grade baselines that reflect fundamentally different architectural philosophies:
\begin{itemize}[leftmargin=*,itemsep=2pt]
    \item \textbf{Wukong}~\citep{zhang2024wukong}: Establishes principled scaling laws for cross-feature modeling through stacked factorization-machine based interaction modules, providing foundational insights for scalable non-sequential recommendation.
    \item \textbf{InterFormer}~\citep{zeng2024interformer}: Bridges sequence and feature modeling through learnable interaction tokens with bidirectional information flow, representing the current production state-of-the-art for joint sequence-nonsequence modeling.
\end{itemize}

Our baselines are chosen for \textit{production readiness} (both serve billions of daily requests at Meta), \textit{scale equivalence} (identical datasets, feature vocabularies, and computational budgets), and \textit{paradigm coverage} (representing complementary architectural approaches---pure feature interaction vs. joint modeling).

We did not include HSTU~\citep{zhai2024actions} in our comparison because it is specifically designed for sequence-centric and request-level optimization scenarios. In contrast, our experiments operate at the impression level, utilizing hundreds to thousands of non-sequential contextual features that are critical for CTR prediction in production systems. Additionally, non-sequential context features continue to play a very important role in performance. If we do not handle and model them appropriately, it could result in significant performance regressions. This fundamental mismatch in design assumptions (i.e., HSTU's focus on behavior-dominated scenarios vs. our heterogeneous feature setting) makes a direct comparison less meaningful. While HSTU demonstrates strong scaling in its target scenarios, our work addresses the challenge of scaling in settings where both sequential and non-sequential features are present.

\noindent\textbf{Evaluation Metrics.} We use the following metrics:
\begin{itemize}
\item \textbf{NE}: Primary offline metric for CTR prediction quality, lower is better (Equation~\ref{eq:ne})
\item \textbf{MFU}: Ratio of achieved FLOPs to peak hardware FLOPs, indicating computational efficiency
\item \textbf{QPS}: Training throughput of the distributed system
\item \textbf{Scaling Efficiency}: NE gain per unit of Total Compute, measuring the rate of performance improvement (Equation~\ref{eq:scaling_eff})
\end{itemize}


\noindent\textbf{Implementation Details.} \algname\ is implemented in PyTorch with custom Triton kernels for performance-critical operations (GDPA, HSP). The model supports training on a variety of hardware setups, including NVIDIA H100, B200, and GB200 GPUs. For fair comparisons in our experiments, we use B200, consistent global batch sizes, and optimizer configurations across all runs, unless otherwise specified. All experiments are trained for a single epoch.

\subsection{Main Results: Architecture Comparison at Different Scales (RQ1)}

\begin{figure}
    \centering
    \includegraphics[width=0.65\linewidth]{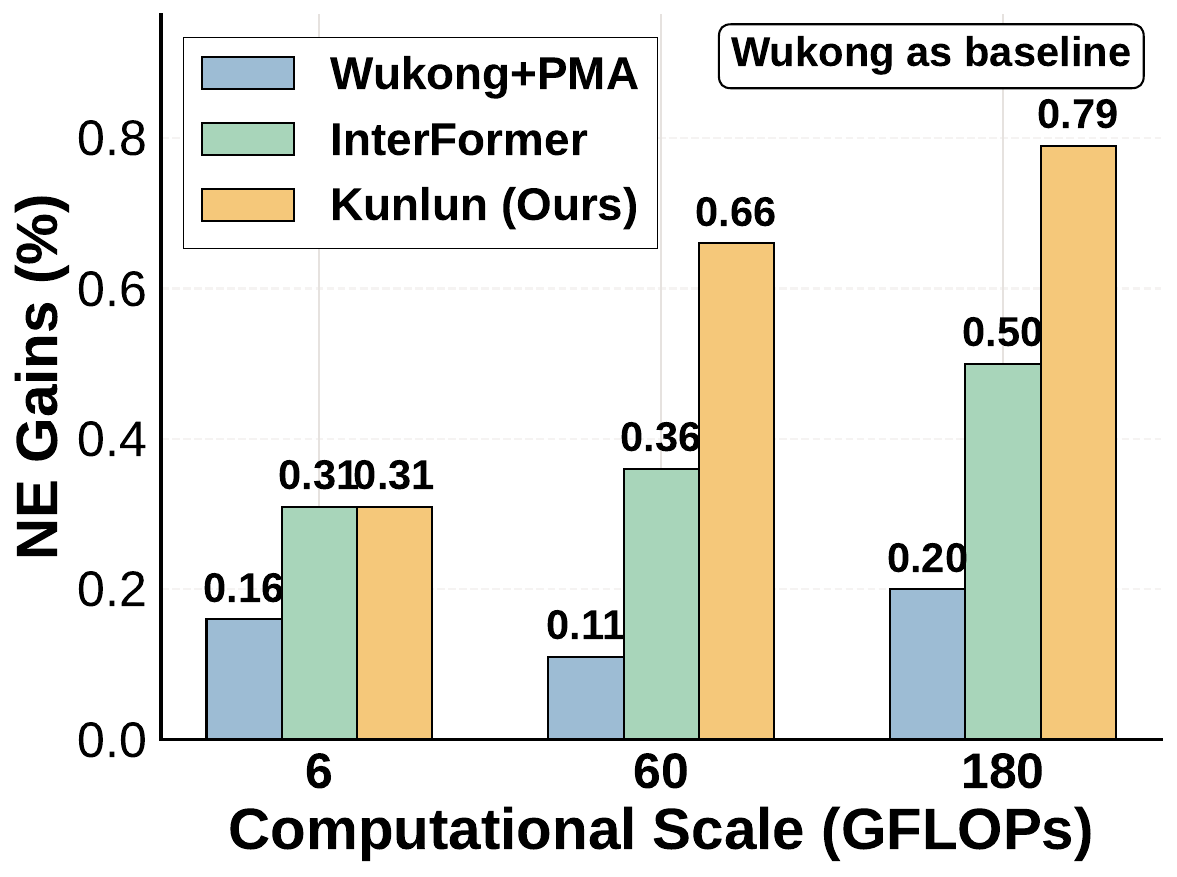}
    \caption{NE gains of different architectures compared to Wukong baseline across computational scales (6, 60, and 180 GFLOPs). The y-axis shows the absolute NE gains, where lower NE indicates better model performance. Larger values represent greater improvement over Wukong. \algname\ achieves the largest NE gains at all scales (0.31\%, 0.66\%, 0.79\%), with the performance gap widening as computational budget increases, demonstrating superior scaling efficiency. Note: NEs are not comparable across scales due to different feature configurations; within-scale results use identical setups.
    }
    \label{fig:architecture_comparison}
\end{figure}

Figure~\ref{fig:architecture_comparison} presents the comparison of different architectures at three computational scales (6 GFLOPs, 60 GFLOPs, and 180 GFLOPs), showing the NE gains over Wukong baseline. \algname\ consistently achieves the largest NE gains across all scales. Notably, while \algname\ and InterFormer show comparable performance at the 6 GFLOPs scale (both achieving 0.31\% NE gains), the performance gap widens significantly at larger scales. As computational budget increases from 6 to 180 GFLOPs, \algname's NE gains grow from 0.31\% to 0.79\% (2.5$\times$), while InterFormer only improves from 0.31\% to 0.50\% (1.6$\times$). This \textbf{superior scaling efficiency} validates our hypothesis that addressing both inefficient modules (low-level) and inefficient computation allocation (high-level) is necessary for achieving predictable scaling laws.


\begin{table*}[t]
\centering
\setlength{\tabcolsep}{10pt}
\caption{Ablation study on \algname\ components. We report the impact of removing each component on MFU, QPS, NE, and GFLOPs. $\Delta$QPS (negative means worse), $\Delta$NE (positive means worse), and $\Delta$GFLOPs are relative to full \algname\ model .}
\label{tab:ablation_study}
\begin{tabular}{llcccc}
\toprule
\textbf{Category} & \textbf{Configuration} & \textbf{MFU (\%)} & \textbf{$\Delta$QPS (\%)} & \textbf{$\Delta$NE (\%)} & \textbf{$\Delta$GFLOPs} \\
\midrule
\multirow{4}{*}{\textit{\makecell[l]{Low-Level (RQ2)}}} 
& w/o GDPA & 34.0\% & -8\% & +0.04\% & Neutral \\
& w/o HSP (use PMA) & 30.21\% & +8.8\% & +0.08\% & -25\% \\
& w/o SumKronLinear & 34.55\% & Neutral & +0.03\% & -7.5\% \\
& w/o Sliding Window & 32.90\% & -31.1\% & Neutral & +29.5\% \\
\midrule
\multirow{3}{*}{\textit{\makecell[l]{High-Level (RQ3)}}} 
& w/o CompSkip & 34.50\% & -35\% & Neutral & +43.1\% \\
& w/o Event Personalization & 35.7\% & -13\% & -0.02\% & +11\% \\
& w/o Expert Parallelism & 35.5\% & -4\% & Neutral & Neutral \\
\midrule
\textit{Full Model} & \textbf{\algname\ (Full)} & \textbf{37.0\%} & \textbf{Baseline} & \textbf{Baseline} & \textbf{154.9} \\
\bottomrule
\end{tabular}
\vspace{0.1cm}
\\
\footnotesize{MFU measured on NVIDIA B200 GPUs. All experiments use identical hardware and training configurations.}
\end{table*}

\subsection{Ablation Study: Component Analysis (RQ2 \& RQ3)}

Table~\ref{tab:ablation_study} presents the ablation study analyzing the contribution of each component in \algname.

\noindent\textbf{Low-Level Module Optimizations (RQ2).} The GDPA reformulation contributes to overall MFU improvement: removing GDPA reduces MFU from 37.0\% to 34.0\% and causes 8\% QPS degradation, validating that the attention-style reformulation enables more efficient GPU utilization. Without GDPA, NE degrades by 0.04\%, confirming its contribution to both efficiency and model quality. HSP improves model quality over traditional PMA, achieving 0.08\% NE gain. While PMA offers 8.8\% higher QPS due to its simpler design, HSP's superior sequence summarization quality justifies the trade-off for production systems where NE gains directly impact business metrics. SumKronLinear contributes 0.03\% NE improvement through its expressive cross-dimensional compression while maintaining QPS neutrality. Sliding Window Attention provides substantial efficiency gains, including 31.1\% QPS improvement and 29.5\% FLOPs reduction, while preserving model quality. This confirms the strong locality bias in recommendation sequences.

\noindent\textbf{High-Level Computation Reallocation (RQ3).} CompSkip delivers the largest efficiency gains, achieving 43.1\% FLOPs reduction and 35\% QPS improvement through the every-other-layer pattern while maintaining NE neutrality. This validates layer redundancy in deep recommendation models. Event-level Personalization provides 13\% QPS improvement and 11\% FLOPs reduction by allocating resources proportional to event importance. Expert Parallelism contributes 4\% QPS improvement through distributed computation while maintaining both NE and FLOPs neutrality, demonstrating effective communication-computation overlap.

\subsection{Scaling Law Analysis}

\textbf{Scaling Efficiency Comparison.} \algname\ achieves \textbf{2$\times$ scaling efficiency over state-of-the-art approaches}---doubling the rate at which performance improves with additional compute compared to InterFormer.

\begin{figure}[t]
\centering
\includegraphics[width=.49\linewidth]{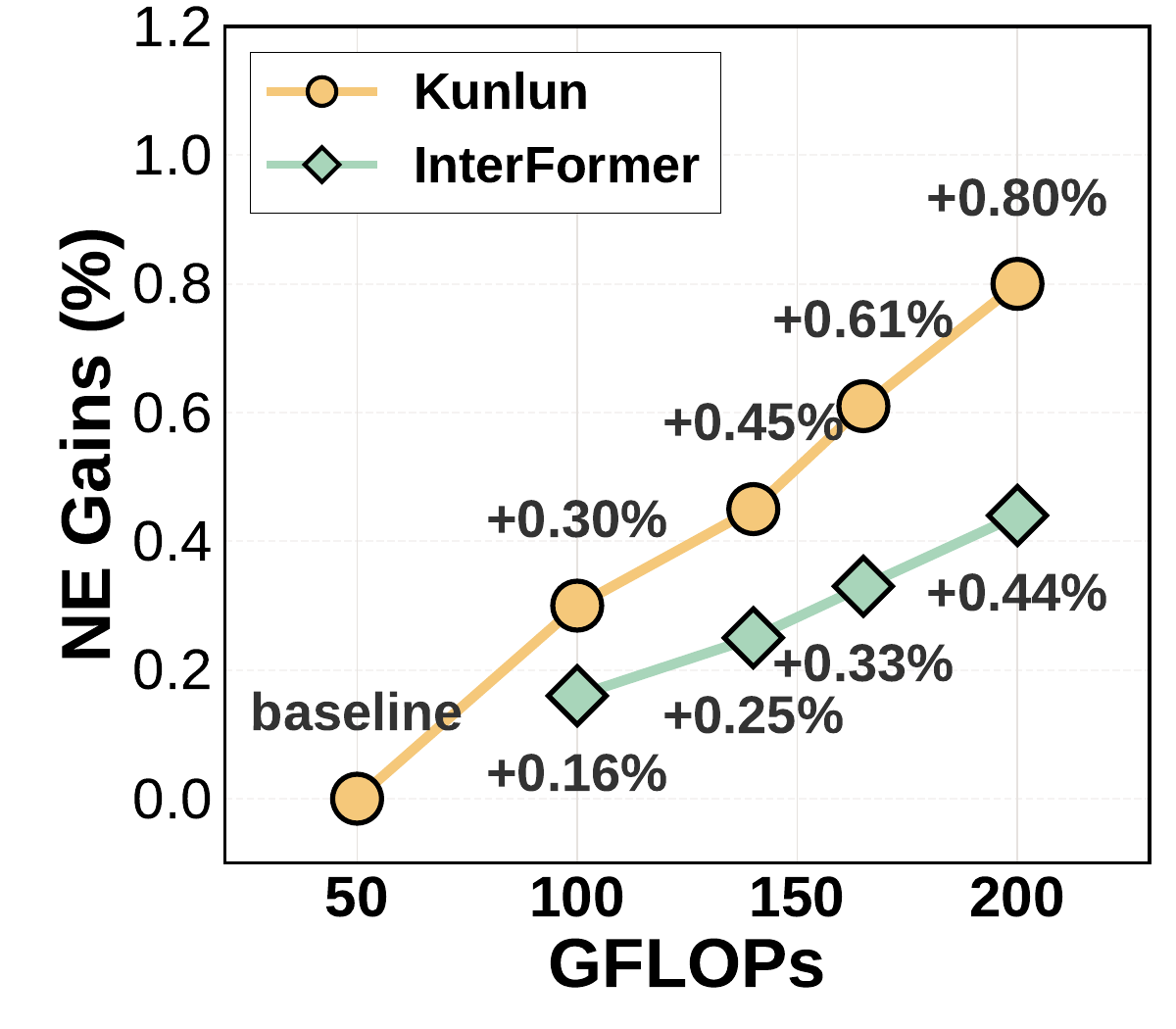}
\includegraphics[width=.49\linewidth]{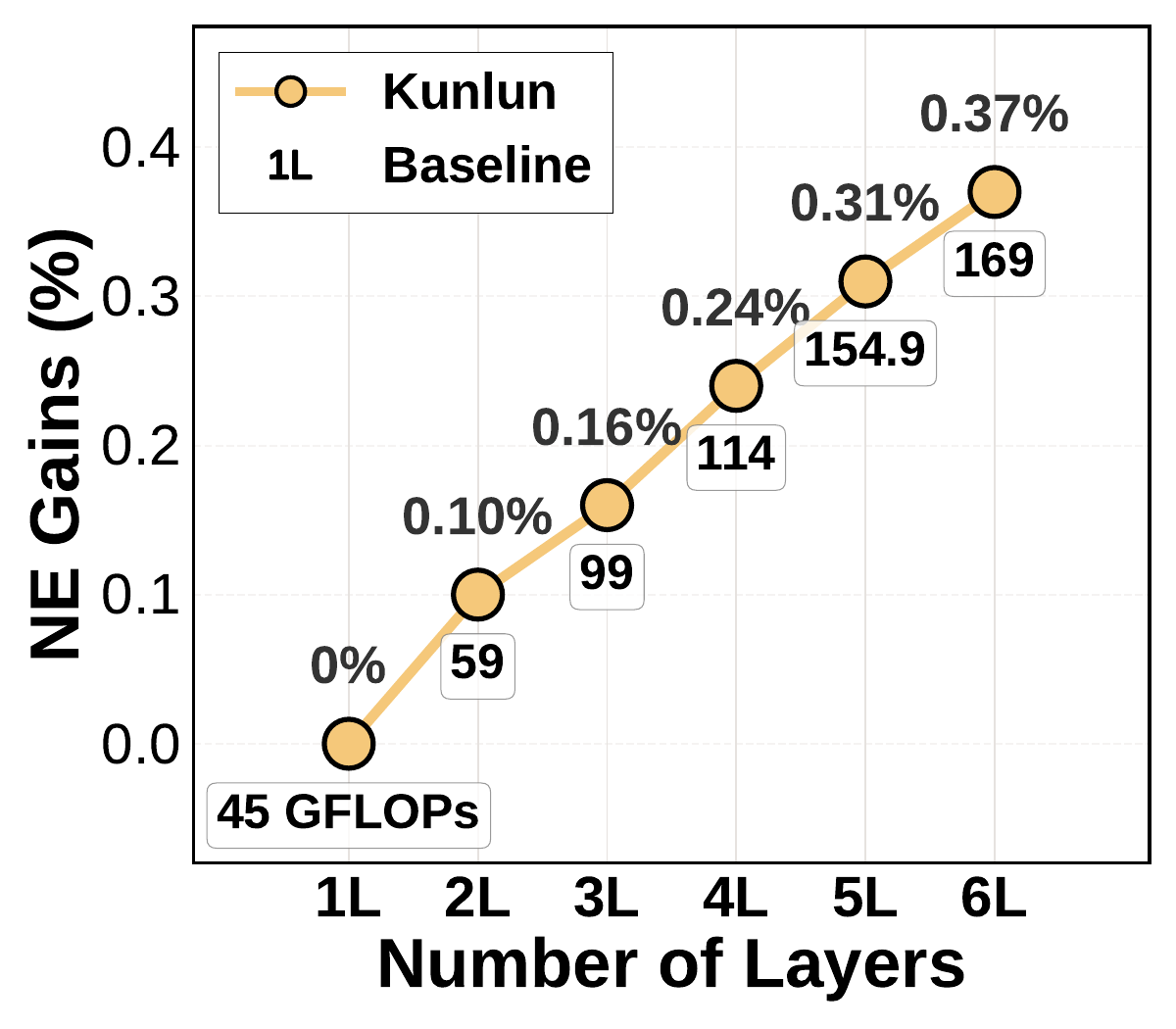}
\vspace{-0.6cm}
\caption{\label{fig:scaling_curves}(Left) Scaling law curves showing NE improvement vs. Total Compute. \algname\ achieves steeper scaling (2$\times$ efficiency over state-of-the-art) with predictable power-law behavior. \label{fig:layer_scaling}(Right) NE improvement as a function of number of layers (1-6). Each additional layer provides diminishing but predictable returns following logarithmic scaling.}
\end{figure}

\noindent\textbf{Scaling Law Curves.} Figure~\ref{fig:scaling_curves} (left) shows NE improvement as a function of Total Compute (GFLOPs) for different architectures. \algname\ exhibits predictable power-law scaling behavior consistent with Equation~\ref{eq:scaling_law}, with the scaling coefficient approximately 2$\times$ larger than InterFormer. This predictability enables reliable extrapolation for resource planning and architectural decisions.

\noindent\textbf{Model Scaling.} We evaluate vertical scaling by varying the number of layers from 1 to 6. Each additional layer provides consistent NE gains following the logarithmic pattern in Equation~\ref{eq:scaling_law}, validating our multi-layer architecture design. Specifically, we observe that NE improvement from layer $l$ to $l+1$ follows $\Delta \text{NE}_l \approx c / \log(l+1)$, confirming diminishing but predictable returns (see Figure~\ref{fig:scaling_curves} (right)).

\subsection{Production Deployment Results (RQ4)}

\algname\ has been deployed across major Meta Ads models, delivering a 1.2\% improvement in topline metrics and demonstrating superior ROI at scale compared to state-of-the-art architectures.

\section{Conclusion}\label{sec:con}

We present \algname, a unified architecture that establishes predictable scaling laws for massive-scale recommendation systems through systematic model-efficiency codesign. By addressing computational inefficiency through low-level module optimization (GDPA, HSP, Sliding Window Attention) and suboptimal computation allocation through high-level reallocation (CompSkip, Event Personalization, Expert Parallelism), \algname\ achieves superior NE gains and 2X scaling efficiency over state-of-the-art approaches. The architecture demonstrates strong generalization across diverse recommendation scenarios and has been successfully deployed at major Meta Ads models with significant online impact. Our work provides a principled framework for scaling recommendation systems to unprecedented scales while maintaining computational efficiency, opening new possibilities for future exploration in unified modeling of heterogeneous features.

\clearpage
\newpage
\balance
\bibliographystyle{assets/plainnat}
\bibliography{reference}

\clearpage
\newpage
\beginappendix

\appendix

\section{Evaluation Metrics Details}\label{app:metrics}

\subsection{Normalized Entropy (NE)}

NE is defined as the ratio of model cross-entropy loss to background entropy:
\begin{equation}\label{eq:ne}
\text{NE} = \frac{\mathcal{L}(y, \hat{y})}{\mathcal{H}(p)}
\end{equation}
where the cross-entropy loss is:
\begin{equation}\label{eq:ce_loss_full}
\mathcal{L}(y, \hat{y}) = -\frac{1}{N} \sum_{i=1}^{N} \left[ y_i \log(\hat{y}_i) + (1 - y_i) \log(1 - \hat{y}_i) \right]
\end{equation}
and the background entropy $\mathcal{H}(p) = -p \log(p) - (1-p) \log(1-p)$ represents the entropy of predicting using only the empirical CTR $p = \frac{1}{N} \sum_{i=1}^{N} y_i$.

\textbf{Properties.} (1) \textit{Range}: NE $\in [0, +\infty)$, with perfect models achieving NE $= 0$ and baseline models achieving NE $= 1$. (2) \textit{Data-Insensitivity}: Normalization by $\mathcal{H}(p)$ enables comparison across datasets with different CTR distributions. (3) \textit{Calibration Sensitivity}: Unlike AUC, NE penalizes miscalibrated predictions.

\textbf{Why NE over AUC?} In production systems, calibration is critical because predicted probabilities are used directly in auction mechanisms, budget pacing, and revenue optimization.

\subsection{Model FLOPs Utilization (MFU)}

MFU measures hardware utilization efficiency:
\begin{equation}\label{eq:mfu_full}
\text{MFU} = \frac{\text{Model FLOPs} \times \text{Throughput}}{\text{GPU Peak TFLOPS} \times \text{Num GPUs}}
\end{equation}
Low MFU ($<$20\%) indicates memory-bound operations; high MFU ($>$40\%) indicates compute-bound operations that effectively utilize GPU resources. In this paper, MFU is measured on a single GPU and reflects only the model architecture's intrinsic GPU efficiency,independent of system-level overheads.

\section{Feature Preprocessing Details}\label{app:preprocessing}

Before entering the first layer, the raw features are transformed into unified embeddings that the model can process.

\subsection{Non-Sequence Feature Preprocessing}

Non-sequence features consist of two types:

\textbf{Dense Features} ($\mathbf{x}_{\text{dense}}^{(0)}$): Continuous values such as user age, item price, and temporal context (time-of-day, day-of-week). Raw dense features $x_{\text{dense}_i}^{(0)}$ are concatenated to form a dense vector $\mathbf{x}_{\text{dense}}^{(0)}=\left[x_{\text{dense}_1}^{(0)},\dots,x_{\text{dense}_m}^{(0)}\right]^\top \in \mathbb{R}^{m}$, which is transformed into a $d$-dimensional embedding via linear projection:
\begin{equation}\label{eq:dense_preproc}
    \mathbf{x}_{\text{dense}}^{(1)}=\mathbf{W}_{\text{dense}}\mathbf{x}_{\text{dense}}^{(0)} \in \mathbb{R}^{d},
\end{equation}
where $\mathbf{W}_{\text{dense}} \in \mathbb{R}^{d \times m}$ is a learnable projection matrix and the superscript $(1)$ denotes the output after preprocessing (layer 1 input).

\textbf{Sparse Features} ($\mathbf{x}_{\text{sparse}_i}^{(0)}$): Categorical values such as user id, item category, and location. Each sparse feature is first encoded as a one-hot vector $\mathbf{x}_{\text{sparse}_i}^{(0)}\in\mathbb{R}^{n_{v_i}}$, where $n_{v_i}$ is the vocabulary size of the $i$-th sparse feature, then embedded via:
\begin{equation}\label{eq:sparse_preproc}
    \mathbf{x}_{\text{sparse}_i}^{(1)}=\mathbf{W}_{\text{sparse}_i}\mathbf{x}_{\text{sparse}_i}^{(0)} \in \mathbb{R}^{d},
\end{equation}
where $\mathbf{W}_{\text{sparse}_i} \in \mathbb{R}^{d \times n_{v_i}}$ is a learnable embedding matrix.

By concatenating the dense and sparse embeddings, we obtain the unified non-sequence representation:
\begin{equation}\label{eq:nonseq_concat}
    \mathbf{X}^{(1)}=\left[\mathbf{x}_{\text{dense}}^{(1)};\mathbf{x}_{\text{sparse}_1}^{(1)};\dots;\mathbf{x}_{\text{sparse}_n}^{(1)}\right] \in \mathbb{R}^{(n+1) \times d},
\end{equation}
where $;$ denotes concatenation along the feature dimension. This unified representation contains $(n+1)$ features, each represented as a $d$-dimensional vector.

\subsection{Sequence Feature Preprocessing}

Sequence features capture dynamic user interests through interaction history. Each interacted item in the sequence is embedded as a $d$-dimensional vector $\mathbf{s}_{t}^{(0)}$, and sequences are represented as:
\begin{equation}\label{eq:seq_basic}
    \mathbf{S}^{(0)}=\left[\mathbf{s}_1^{(0)};\dots;\mathbf{s}_T^{(0)}\right]\in\mathbb{R}^{T \times d},
\end{equation}
where $T$ is the sequence length.

\textbf{Multi-Sequence Integration.} Real-world systems often contain multiple sequences from different event types (clicks, conversions, impressions) or platforms. To unify these heterogeneous sequences and filter internal noise, we employ an MLP. Given $k$ sequences $\mathbf{S}_1^{(0)},\cdots,\mathbf{S}_k^{(0)}$, they are first concatenated along the embedding dimension to form $\left[\mathbf{S}_1^{(0)};\cdots;\mathbf{S}_k^{(0)}\right]\in\mathbb{R}^{T \times kd}$, then processed via:
\begin{equation}\label{eq:masknet}
\text{MLP}_{\text{lce}}(\mathbf{S}),
\end{equation}
where $\text{MLP}_{\text{lce}}: \mathbb{R}^{T \times kd}\to \mathbb{R}^{T \times d}$ linearly combines multiple sequences into a single unified representation matching the dimensionality of non-sequence features.

\textbf{Rotary Temporal Embeddings (ROTE).} A key innovation in \algname\ is extending Rotary Position Embeddings (RoPE)~\citep{su2024roformer} to Rotary Temporal Embeddings (ROTE) for recommendation sequences. Unlike LLMs where token position is the primary ordering signal, recommendation sequences exhibit temporal patterns where the \textit{time gap} between events carries crucial information---a click from yesterday is fundamentally different from a click from last month, even if they occupy adjacent positions. ROTE computes timestamp gaps between consecutive events and applies log-scale encoding to handle the wide range of temporal intervals (from seconds to months). These log-transformed timestamp gaps are incorporated into the rotary embeddings, enabling the model to distinguish not only the sequential order of interactions but also their temporal proximity---critical for understanding whether behaviors represent recent interests or historical patterns.

Specifically, for position $t$ with timestamp gap $\Delta t$, we apply log-scale encoding $\tau_t = \log(1 + \Delta t / \tau_{\text{scale}})$ and incorporate it into rotary embeddings:
\begin{equation}
\text{ROTE}(\mathbf{x}, t, \tau_t) = \mathbf{R}_{t,\tau_t} \mathbf{x}
\end{equation}
where $\mathbf{R}_{t,\tau_t}$ is a block-diagonal rotation matrix with elements $\cos(t\theta_i + \tau_t\phi_i)$ and $\sin(t\theta_i + \tau_t\phi_i)$ for frequency parameters $\theta_i, \phi_i$.

\section{Algorithm Details}\label{app:algorithm}
\subsection{Overview}
This section provides the complete algorithmic descriptions of the key components in \algname. We present four algorithms: (1) the main layer forward pass, (2) Hierarchical Seed Pooling (HSP), (3) Generalized Dot-Product Attention (GDPA), and (4) CompSkip configuration generation.

\subsection{Kunlun Forward Pass}
Algorithm~\ref{alg:kunlun_layer} describes the forward pass for a single \algname\ layer. Each layer processes both non-sequence features $\mathbf{X}^{(l)}$ and sequence features $\mathbf{S}^{(l)}$, producing refined representations for the next layer. The algorithm incorporates CompSkip logic, which conditionally skips certain computations (HSP, PFFN, or self-attention) based on the layer configuration to reduce computational cost while maintaining model quality.

\begin{algorithm}[h]
\caption{\algname\ Layer Forward}\label{alg:kunlun_layer}
\begin{algorithmic}[1]
\REQUIRE $\mathbf{X}^{(l)}$, $\mathbf{S}^{(l)}$, $\mathbf{H}_{\text{prev}}$, layer config
\ENSURE $\mathbf{X}^{(l+1)}$, $\mathbf{S}^{(l+1)}$, $\mathbf{H}_{\text{curr}}$
\STATE \textbf{// Kunlun Interaction Block}
\STATE $\mathbf{W}^{(l)} \gets \text{WeightGen}(\text{Summarize}(\mathbf{X}^{(l)}))$
\IF{not skip\_hsp}
    \STATE $\mathbf{H}_{\text{curr}} \gets \text{HSP}(\mathbf{S}^{(l)})$ \COMMENT{Alg.~\ref{alg:hsp}}
\ELSE
    \STATE $\mathbf{H}_{\text{curr}} \gets \mathbf{H}_{\text{prev}}$
\ENDIF
\STATE $\mathbf{X}^{(l+1)} \gets \text{GlobalInteraction}([\mathbf{X}^{(l)} | \mathbf{H}_{\text{curr}}])$
\STATE \textbf{// Kunlun Transformer Block}
\IF{not skip\_pffn}
    \STATE $\tilde{\mathbf{S}} \gets \text{GDPA}(\mathbf{S}^{(l)}, \mathbf{W}^{(l)}) + \mathbf{S}^{(l)}$ \COMMENT{Alg.~\ref{alg:gdpa}}
\ELSE
    \STATE $\tilde{\mathbf{S}} \gets \mathbf{S}^{(l)}$
\ENDIF
\IF{not skip\_self\_attention}
    \STATE $\mathbf{S}^{(l+1)} \gets \text{SlidingWindowMHA}(\tilde{\mathbf{S}}) + \tilde{\mathbf{S}}$
\ELSE
    \STATE $\mathbf{S}^{(l+1)} \gets \tilde{\mathbf{S}}$
\ENDIF
\RETURN $\mathbf{X}^{(l+1)}$, $\mathbf{S}^{(l+1)}$, $\mathbf{H}_{\text{curr}}$
\end{algorithmic}
\end{algorithm}

\begin{algorithm}[h]
\caption{Hierarchical Seed Pooling (HSP)}\label{alg:hsp}
\begin{algorithmic}[1]
\REQUIRE $\mathbf{S} \in \mathbb{R}^{B \times T \times d}$, seeds $n_s$, tokens $n_t$
\ENSURE $\mathbf{H} \in \mathbb{R}^{B \times n_t \times d}$
\STATE $\mathbf{E} \gets \text{LearnableSeeds}(n_s, d)$
\STATE $\mathbf{H}_s \gets \text{MHA}(\text{Norm}(\mathbf{E}), \mathbf{S}, \mathbf{S})$ \COMMENT{Seed attention}
\STATE $\mathbf{H} \gets \sum_{i=1}^{k} \mathbf{Z}_i^\top \mathbf{H}_s \mathbf{W}_i$ \COMMENT{SumKronLinear}
\RETURN $\mathbf{H}$
\end{algorithmic}
\end{algorithm}

\subsection{Hierarchical Seed Pooling (HSP)}

Algorithm~\ref{alg:hsp} describes the Hierarchical Seed Pooling mechanism for sequence summarization. HSP compresses long sequences into compact representations through three stages: (1) initializing learnable seed embeddings that are shared across the batch, (2) computing seed-level attention where seeds attend to the full sequence, and (3) applying SumKronLinear compression to reduce seeds to the final output tokens. This hierarchical approach provides better initialization and more stable training compared to traditional random query-based pooling methods.

\subsection{Generalized Dot-Product Attention (GDPA)}

Algorithm~\ref{alg:gdpa} presents the GDPA mechanism that reformulates the traditional Personalized FeedForward Network (PFFN) as an attention-like operation. The key insight is that sequence transformation guided by non-sequence context can be viewed as cross-attention, where sequence tokens serve as queries and the non-sequence context generates keys and values through learnable projections. Each head can use a different activation function (Act$_h$), providing ensemble diversity. This reformulation converts memory-bound operations into compute-bound matrix multiplications, significantly improving MFU.

\begin{algorithm}[h]
\caption{Generalized Dot-Product Attention (GDPA)}\label{alg:gdpa}
\begin{algorithmic}[1]
\REQUIRE $\mathbf{S} \in \mathbb{R}^{B \times T \times d}$, $\mathbf{X}_{\text{sum}} \in \mathbb{R}^{B \times n \times d}$
\ENSURE $\mathbf{S}' \in \mathbb{R}^{B \times T \times d}$
\FOR{$h = 1$ \TO $H$}
    \STATE $\mathbf{K}_h, \mathbf{V}_h \gets \mathbf{w}_1^{(h)}(\mathbf{X}_{\text{sum}}), \mathbf{w}_2^{(h)}(\mathbf{X}_{\text{sum}})$
    \STATE $\mathbf{Q}_h \gets \mathbf{W}_h^Q \mathbf{S}$
    \STATE $\mathbf{O}_h \gets \text{Act}_h(\mathbf{Q}_h \mathbf{K}_h^\top / \tau) \mathbf{V}_h$
\ENDFOR
\STATE $\mathbf{S}' \gets \mathbf{W}^O [\mathbf{O}_1 | \dots | \mathbf{O}_H]$
\RETURN $\mathbf{S}'$
\end{algorithmic}
\end{algorithm}

\subsection{CompSkip Configuration}

Algorithm~\ref{alg:compskip} generates the CompSkip configuration for a model with $L$ layers. The every-other-layer pattern alternates between two configurations: even layers (0, 2, 4, ...) skip self-attention but compute fresh HSP summaries, while odd layers (1, 3, 5, ...) skip HSP (reusing summaries from the previous layer) and PFFN but compute self-attention. This design ensures that each layer contributes either local refinement (self-attention) or global summarization (HSP), maintaining model expressiveness while significantly reducing computation.


\begin{algorithm}[h]
\caption{CompSkip Configuration}
\label{alg:compskip}
\begin{algorithmic}[1]
\REQUIRE Number of layers $L$
\ENSURE Layer configurations
\FOR{$l = 0$ \TO $L-1$}
    \IF{$l$ is even}
        \STATE skip\_self\_attention $\gets$ True
        \STATE skip\_hsp $\gets$ False
    \ELSE
        \STATE skip\_self\_attention $\gets$ False
        \STATE skip\_hsp $\gets$ True
        \STATE skip\_pffn $\gets$ True
    \ENDIF
\ENDFOR
\end{algorithmic}
\end{algorithm}

\subsection{Complexity Analysis}

Table~\ref{tab:complexity} summarizes the computational complexity of each component.

\begin{table}[h]
\centering
\caption{Time complexity per layer. $T$: sequence length, $N$: non-sequence features, $d$: embedding dim, $w$: window size, $M$: experts.}
\label{tab:complexity}
\small
\begin{tabular}{lc}
\toprule
\textbf{Component} & \textbf{Complexity} \\
\midrule
GDPA & $O(T \cdot N \cdot d)$ \\
Sliding Window Attention & $O(T \cdot w \cdot d)$ \\
Full Self-Attention & $O(T^2 \cdot d)$ \\
HSP & $O(n_s \cdot T \cdot d)$ \\
Global Interaction & $O(M \cdot (N + n_t)^2 \cdot d)$ \\
\bottomrule
\end{tabular}
\end{table}

\textbf{CompSkip Savings.} The every-other-layer pattern reduces self-attention, HSP, and PFFN FLOPs each by $\sim$50\%, achieving 43.1\% total FLOPs reduction.

\section{Multi-Layer Architecture and Information Flow Details}\label{app:multilayer}
\textbf{Layer Stacking.} \algname\ stacks $L$ layers, where each layer $l \in \{0, \dots, L-1\}$ consists of two main blocks that process sequential and non-sequential features through systematic interaction:

\textit{Kunlun Transformer Block.} Context-aware sequence modeling:
\begin{equation}\label{eq:kunlun_transformer}
\mathbf{S}^{(l+1)} = \text{MHA}(\text{GDPA}(\mathbf{W}^{(l)}, \mathbf{S}^{(l)})),
\end{equation}
where GDPA-enhanced PFFN first performs personalized sequence transformation using personalized weights $\mathbf{W}^{(l)}$ generated from non-sequence features, followed by Multi-Head Self-Attention (MHA) to capture dependencies within the sequence itself. This block outputs refined sequence representations $\mathbf{S}^{(l+1)}$.

\textit{Kunlun Interaction Block.} Cross-modal interaction learning through three modules:

First, the Weight Generation module derives personalized transformation weights from non-sequence features for the next layer's GDPA:
\begin{equation}\label{eq:weight_gen}
\mathbf{W}^{(l+1)} = \text{WeightGen}(\mathbf{X}_{\text{sum}}^{(l)}),
\end{equation}
where $\mathbf{X}_{\text{sum}}^{(l)}$ are summarized non-sequence features. While this module feeds into the Kunlun Transformer Block, it is conceptually part of the Kunlun Interaction Block as it represents information flow from non-sequence to sequence features---a key bidirectional interaction mechanism.

Second, the HSP module produces compact sequence summaries:
\begin{equation}\label{eq:seq_sum}
[\mathbf{S}^{(l+1)}_{\text{CLS}}, \mathbf{S}^{(l+1)}_{\text{HSP}}, \mathbf{S}^{(l+1)}_{\text{recent}}] = \text{HSP}(\mathbf{S}^{(l+1)}),
\end{equation}
where three types of summaries are produced: CLS tokens learned through attention, HSP tokens through hierarchical seed-based compression, and recent tokens capturing immediate temporal context.

Third, the Global Interaction module processes both non-sequence features and sequence summaries through a mixture of Wukong experts:
\begin{equation}\label{eq:global_interaction}
\mathbf{X}^{(l+1)} = \text{GlobalInteraction}([\mathbf{X}^{(l)} | \mathbf{S}^{(l+1)}_{\text{CLS}} | \mathbf{S}^{(l+1)}_{\text{HSP}} | \mathbf{S}^{(l+1)}_{\text{recent}}]),
\end{equation}
where sequence summaries are concatenated with non-sequence features, and multiple Wukong experts process different feature partitions concurrently to learn rich cross-modal interactions. The summarized non-sequence output $\mathbf{X}^{(l+1)}_{\text{sum}}$ is then used by the Weight Generation module for the next layer.

\textbf{Bidirectional Information Flow.} The architecture maintains bidirectional information exchange between sequential and non-sequential modalities, extending the interaction paradigm established in prior joint modeling work~\citep{zeng2024interformer}: (1) non-sequence features guide sequence modeling through the Weight Generation module that produces personalized weights for GDPA in the Kunlun Transformer Block (Eq.\ref{eq:weight_gen} and \ref{eq:kunlun_transformer}), and (2) sequence summaries inform non-sequence interaction through concatenation in the Global Interaction module (Eq.\ref{eq:global_interaction}). This enables mutually beneficial learning where both modalities progressively refine each other across layers. \algname\ maintains these bidirectional properties while amplifying their effectiveness through the efficiency optimizations described in Sections~\ref{sec:low_level} and~\ref{sec:high_level}.

\textbf{Hierarchical Learning.} Vertical layer stacking enables hierarchical feature learning at progressively higher levels of abstraction, a property also observed in prior work~\citep{zeng2024interformer} that \algname\ preserves and extends: Layer 0 learns basic patterns (e.g., single feature importance and local sequence dependencies), Layer 1 captures pairwise interactions (e.g., user-item affinity conditioned on recent behavior), and deeper layers model high-order dependencies (e.g., complex sequential patterns conditioned on global user context and multi-feature interactions). Each layer operates on increasingly refined representations from the previous layer, with the Kunlun Transformer Block extracting richer sequence representations and the Kunlun Interaction Block learning more sophisticated cross-modal interactions.

\textbf{Scaling Properties.} The architecture demonstrates predictable scaling behavior along both vertical (depth) and horizontal (width) dimensions. Adding more layers consistently improves NE with diminishing returns following a logarithmic pattern, enabling optimal layer depth selection based on compute budget. Specifically, we observe that NE improvement from layer $l$ to $l+1$ follows $\Delta \text{NE}_l \approx c / \log(l+1)$ where $c$ is a constant, validating the power-law relationship described in Section~\ref{sec:scaling_laws}. Similarly, increasing the number of Wukong experts within each Global Interaction module (horizontal scaling) provides predictable performance gains that follow power-law scaling.

\textbf{Final Prediction.} After $L$ layers, the final non-sequence representation $\mathbf{X}^{(L)}$ from the Global Interaction module is fed into task-specific heads for CTR prediction:
\begin{equation}\label{eq:final_pred}
\hat{y} = \sigma(\text{MLP}(\mathbf{X}^{(L)})),
\end{equation}
where MLP is a multi-layer perceptron and $\sigma$ is the sigmoid function producing click probability. The stacked architecture ensures that $\mathbf{X}^{(L)}$ encodes both rich feature interactions learned through the mixture of Wukong experts and comprehensive sequential patterns captured through bidirectional interaction with the Kunlun Transformer Blocks.

\begin{table*}[t]
\centering
\setlength{\tabcolsep}{4pt} 
\caption{Comparison of different architectures at various computational scales (6 GFLOPs, 60 GFLOPs, 180 GFLOPs baseline compute) with Wukong as the baseline. We report NE Gains (\%) as the absolute NE improvement compared to Wukong, where lower NE indicates better model performance. Larger values indicate greater improvement. \algname\ achieves consistently larger NE gains across all scales with superior scaling efficiency.}
\label{tab:architecture_comparison}
\begin{tabular}{lcccccc}
\toprule
\multirow{2}{*}{\textbf{Method}} & \multicolumn{2}{c}{\textbf{6 GFLOPs Scale}} & \multicolumn{2}{c}{\textbf{60 GFLOPs Scale}} & \multicolumn{2}{c}{\textbf{180 GFLOPs Scale}}  \\
\cmidrule(lr){2-3} \cmidrule(lr){4-5} \cmidrule(lr){6-7}
& NE Gains (\%) & GFLOPs & NE Gains (\%) & GFLOPs & NE Gains (\%) & GFLOPs  \\
\midrule
Wukong~\citep{zhang2024wukong} & Baseline & 7.3 & Baseline & 57 & Baseline & 174.5  \\
Wukong + PMA & 0.16\% & 6.1 & 0.11\% & 59.1 & 0.20\% & 187.1  \\
InterFormer~\citep{zeng2024interformer} & 0.31\% & 5.7 & 0.36\% & 57 & 0.50\% & 172.3  \\
\midrule
\textbf{\algname\ (Ours)} & \textbf{0.31\%} & 5.8 & \textbf{0.66\%} & 59.6 & \textbf{0.79\%} & 154.9  \\
\bottomrule
\end{tabular}
\end{table*}

\section{Architecture Comparison Details}\label{app:architecture_comparison}

Table~\ref{tab:architecture_comparison} provides detailed numerical results for the architecture comparison experiments described in Section~\ref{sec:exp}. We evaluate four architectures at three computational scales (approximately 6, 60, and 180 GFLOPs per sample) to assess scaling behavior. All experiments use identical training configurations, including the same dataset, batch size, and optimizer settings.

\textbf{Baselines.} Wukong~\citep{zhang2024wukong} serves as the baseline, representing scalable non-sequential feature interaction. Wukong + PMA extends Wukong with Pooling by Multihead Attention for basic sequence summarization. InterFormer~\citep{zeng2024interformer} represents the state-of-the-art for joint sequence-nonsequence modeling with bidirectional information flow.

\textbf{Key Observations.} At small scale (6 GFLOPs), \algname\ and InterFormer achieve comparable NE gains (0.31\%) over Wukong, both significantly outperforming Wukong + PMA (0.16\%). However, as computational budget increases, \algname\ demonstrates superior scaling: at 180 GFLOPs, \algname\ achieves 0.79\% NE gain compared to InterFormer's 0.50\%, representing a 1.58$\times$ advantage. Notably, \algname\ achieves this superior performance with lower FLOPs (154.9 vs. 172.3 GFLOPs at the 180 GFLOPs scale), demonstrating both better algorithmic effectiveness and computational efficiency.

\section{GDPA and Fused PFFN Kernel}\label{app:gdpa}






The GDPA reformulation exposes a unified attention-style computation pattern shared by multiple interaction modules used in production recommender systems, including self-attention, PMA, and PFFN.
This unified view enables the design of a single, fully fused kernel that replaces the fragmented two-layer MLP and intermediate activations in PFFN.
Inspired by recent FlashAttention kernels, our implementation adopts a memory-efficient, block-wise execution strategy that avoids materializing large intermediate attention tensors in both the forward and backward passes.

Unlike LLM-oriented attention kernels, which are typically optimized for long and dense sequences, GDPA kernels must efficiently handle real-world recommender workloads characterized by short and asymmetric key/value dimensions, large batch sizes, and jagged inputs with variable sequence lengths.
These properties significantly reduce pipeline occupancy and limit compute--memory overlap when using existing attention kernels.
To address this mismatch, we redesign the kernel pipeline by removing softmax-specific stages, simplifying warp specialization, and folding epilogue operations into the activation stage, thereby reducing warp count and freeing register resources for computation-critical warps.

Short K/V sequences pose an additional challenge for persistent attention kernels, as inner-loop software pipelining becomes ineffective when the loop executes only a few iterations.
We therefore restructure the loop nest and apply software pipelining at the outer-loop level, enabling overlap across iterations even when K/V lengths are small.
For jagged inputs, we further introduce software-level tile scheduling based on precomputed sequence-length metadata, eliminating empty tiles and mitigating persistent SM imbalance caused by highly variable workloads.
This scheduling strategy is applied consistently to both forward and backward passes, with appropriate adaptations to their respective loop structures.

Finally, we optimize activation computation to better match GPU execution characteristics.
Since GDPA replaces softmax with element-wise activations, kernels can become SFU-bound under production workloads.
We alleviate this bottleneck by rebalancing transcendental computation toward ALU-dominated approximations and by tightly integrating activation evaluation into the fused kernel pipeline.
Together, these design choices enable a fully fused GDPA-based PFFN kernel that sustains high utilization under irregular, real-world training workloads.

\end{document}